\documentclass{iaus}
\usepackage{graphicx}
\usepackage{psfig}
\usepackage{epsf}

\title[Supernova feedback on star formation] %% give here short title %%
{Supernova Feedback on the Interstellar Medium and Star Formation }

\author[Gerhard Hensler]  %% give here short author list %%
{Gerhard Hensler
}

\affiliation{Institute of Astronomy, University of Vienna,
Tuerkenschanzstr. 17, 1180 Vienna, Austria \\
email: gerhard.hensler@univie.ac.at}

\pubyear{2010}
\volume{270}  %% insert here IAU Symposium No.
\pagerange{119--126}
\date{?? and in revised form ??}
\jname{Computational Star Formation}
\editors{J. Alves, B. Elmegreen, \& V. Trimble, eds.}

\begin{document}
\def\HI{H{\sc i} }
\def\HII{H{\sc ii} }
\def\Ha{{\rm H}\alpha }
\def\SSF{\Sigma_{\rm SF} }
\def\SHa{\Sigma_{\Ha} } 
\def\Sg{\Sigma_{\rm g} }
\def\rSF{\rho_{\rm SF} }
\def\rg{\rho_{\rm g} }
\def\Msun{M_{\odot} }
\def\sfr{$\Msun yr^{-1}$}
\def\Mpc2{\Msun/pc^2}
\def\tff{\tau_{\rm ff} }
\def\tSF{\tau_{\rm SF} } 
\def\e{\epsilon} 
\def\eSN{\epsilon_{\rm SN} }
\def\eSF{\epsilon_{\rm SF} }
\def\OH{{12\-+log(O/H)} }
\def\lNO{log(N/O) }
\def\cd{chemo-dynamical }
\def\AA{A{\rm \&}A}

\maketitle

\begin{abstract}
Supernovae are the most energetic stellar events and influence the 
interstellar medium by their gasdynamics and energetics. 
By this, both also affect the star formation positively and negatively.
In this paper, we review the development of the complexity of investigations
aiming at understanding the interchange between supernovae and their 
released hot gas with the star-forming molecular clouds. 
Commencing from analytical studies the paper advances to numerical models
of supernova feedback from superbubble scales to galaxy structure.
We also discuss parametrizations of star-formation and 
supernova-energy transfer efficiencies. Since evolutionary models
from the interstellar medium to galaxies are numerous and apply
multiple recipes of these parameters, only a representative selection
of studies can be discussed here.
\end{abstract}

\firstsection % if your document starts with a section,

\section{Introduction}

Since stars are formed within the coolest molecular material of the 
interstellar medium (ISM), the star-formation rate (SFR) should be 
determined simply by the gas reservoir and by the free-fall time $\tff$
of molecular clouds (\cite{elm02}). This, however, raises a conflict between 
the ISM conditions and observed SFRs in the sense that $\tff$ for 
a typical molecular cloud density of 100 cm$^{-3}$ amounts to 10$^{14}$ sec, 
i.e. $3\cdot 10^6$ yrs. For the total galactic molecular mass of 
$10^9 - 10^{10}\, \Msun$ the SFR should then amount to about
100 to 1000 \sfr, what is by orders of magnitudes larger than observed 
and the gas reservoir within the Milky Way would have been used up today. 
This means, that the SF timescale must be stretched with respect to collapse 
or dynamical timescale by introducing a SF efficiency (SFE) $\eSF$ and its 
definition could read: $\tSF = \eSF^{-1} \cdot \tff$.

Already in 1959 Schmidt argued that the SFR per unit area is related to 
the gas column density $\Sg$ by a power law with exponent $n$. 
Kennicutt (1998) derived from the $\Ha$ luminosity of spiral galaxies
$\SHa$ a vertically integrated and azimuthally averaged SFR 
(in \sfr $pc^{-2}$), i.e. of gas disks in rotational equilibrium,
and found a correlation with $n = 1.4\pm 0.15$ holding over more than 
4 orders of magnitude in $\Sg$ with a drop below a density threshold 
at 10 $\Mpc2$.
While this relation establishes an equilibrium SFR, it is not surprising 
that the slope varies for dynamically triggered SF, as in starburst and 
merger galaxies and for high-z galaxies, when the disks form by gas infall, 
for the latter reaching $n = 1.7 \pm 0.05$ (\cite{bou07}).

The ordinary Kennicutt-Schmidt (KS) relation can be understood by the 
simple analytical assumption allowing for the $\tff - \tSF$ relation 
and for a uniform state of the ISM. Its equilibrium on disk scales 
requires that heating processes counteract to the natural cooling 
of plasmas. 
Besides the possible heating processes from dissipation of dynamical
effects, as there are the differential rotation of the disk, gas infall,
tidal interactions, shocks, etc., to the feedback by freshly produced stars,
not for all of them it is obvious, how effectively they influence
the SF by the expected self-regulation or vice versa trigger it. 

Unfortunately, the issue of a general KS-law 
is confused by the similarity of slopes under various stellar 
feedback strengths (see e.g. sect. 2 in \cite{hen08}). 
\cite{koe95} demonstrated already that the SFR achieves 
a dependence on $\rg^2$, if the stellar heating is compensated by 
collisional-excited cooling radiation (e.g. \cite{bh89}).
The coefficient of this relation determines $\tSF$. Obviously, the SF 
in galaxies, therefore, depends on both the gas content and the energy 
budget of the ISM. 

Since the most efficient stellar energy power is exerted by supernovae (SNe),
and here particularly by the explosions of the shortly living massive stars
as type II SNe, their feedback to the ISM is of fundamental relevance for 
the SF. In this paper, we overview and discuss the regulation of the ISM by
the SF feedback thru SNe and more pronounced by their cumulative effect as 
superbubbles. Although the expression {\it feedback} of SNe also includes
their release of freshly produced elements, here we only focus on the
dynamical and energetic issues and refer the reader interested on the
chemical evolutionary consequences to \cite{hen10}.

\section{Supernova feedback} 

\subsection{Energy release and the Interstellar Medium}

As SN explosions were since long known to expel vehemently 
expanding hot gas, \cite{spi56} predicted this hot ISM phase to expand 
from the galactic disk where it cannot be bound or pressure confined 
to form a hot halo gas. 
Not before the 70's and with the aid of observations which made formerly 
unaccessible spectral ranges (from the FIR/submm to X-rays) available,
the existence a hot gas component within our Milky Way 
was manifested and later also perceived in other disk galaxies. 
Although the ISM in its cool molecular component is conditioned for SF,
hot gas regulates its dynamics as driver of shocks and turbulence 
as well as its energetics by heating thru cooling radiation and 
heat conduction, by this, exerting negative stellar feedback.

After these facts became internalized, a significant amount of 
mostly analytical explorations were dedicated to a first understanding 
of the ISM structure, its temporal behaviour with respect 
to the hot ISM (comprehensively reviewed by \cite{spi90}), to  
volume filling factors and mass fractions of the gas phases. 
For this purpose, randomly distributed and temporally exploding SNe  
were considered with cooling and according expansion (\cite{cio91}) 
and with transitions to/from the warm/cool gas (\cite{MO77})
or as non-dynamically interchanging gas phases (see e.g. \cite{hab81,ike83}). 
In general, the action of SNe on the ISM is multifacetted
(\cite{che77}) and to a significant part affecting the SF.
Because SNe type II happen on much shorter timescales than type Ia, 
all the ISM and SF feedback studies focus mostly on their energy deposit
and dynamical issues.

While these latter studies include arbitrary SN rates with a
description of the temporal size evolution of individual supernova 
remnants (SNRs), the SF dependence on the physical state is not yet
properly treated. 
With a toy model consisting of 6 ISM components and at least 
10 interchange processes \cite{ike84} also included the formation of
stars from giant molecular clouds. Those can only form from 
cool clouds which, on the other hand, are swept-up and condensed 
in SN shells, so that SF and SN explosions together with different 
gas phases and interchange processes form a consistent network. 
As a natural effect of this local consideration the SF can oscillates
with the corresponding timescales. 

First dynamical approaches to the structure evolution of the ISM and gas 
disks, aiming at understanding the excitation of turbulence, were performed
by \cite{ros95}. As heating sources they took the energy of massive stellar 
winds only into account, which is at the lower range of stellar power 
to the ISM because of their very low energy transfer efficiencies 
(\cite{hen07}). Nevertheless, their models already demonstrate the 
compression of gas filaments and the expulsion of gas vertically from 
the disk even with the inclusion of self-gravity.
Indeed, the SF recipe is not self-consistent and the vertical
boundary conditions seem to affect the vertical gas dynamics artificially.

In more recent numerical investigations of the ISM evolution in the
context of SF and SN explosions \cite{sly05} performed studies of the 
influence of SN feedback, self-gravity, and stars as an additional 
gravitational component on the SFR. 
The main issues can be summarized as that feedback enhances the ISM 
porosity, increases the gas velocity dispersion and the contrasts 
of T and $\rho$, so that smaller and more pronounced structures
form. Since the SFR depends on most of these variables, 
more importantly, the SN feedback models reach SFRs by a factor 
of two higher than without feedback. 

At the same time, \cite{avi04} simulated the structure evolution of the
solar vicinity in a box of $1 \times 1 \times 10\, kpc^3$ size and 
identified the Local Bubble and its neighbouring Loop I in their models. 
In addition, filamentary neutral gas structures, resolved down 
to 0.625 pc, become visible and the vertical matter cycle. 
However, because of the lack of self-consistent SF they applied an 
analytical recipe of random SN events (\cite{avi00}) to their models 
that obviously underestimates the production of superbubbles from 
massive OB associations and, moreover, cannot reasonably achieve
issues on the feedback of SNe on the SF. The ISM processes also do
not include self-regulation effects by radiation or heat conduction, 
which lead to negative SF feedback.

\subsection{Star-formation triggering}

SN and stellar wind-driven bubbles sweep up surrounding 
gas, condense it, and are, by this, expected to excite SF in a 
self-propagating manner (abbreviated SPSF) as a positive feedback. 
In a cornerstone paper \cite{ger78} have demonstrated by a parameter 
study, how SNR expansion can lead to structure formation in galactic 
disks by SPSF. Their main aim, however, to reproduce the general
spiral arm structure thru SPSF by means of the coherence of different 
timescales cannot be reached and most galaxies develop patchy 
structures randomly.

%---------------------------
\begin{figure}[ht]
\begin{center}
\includegraphics[width=6cm]{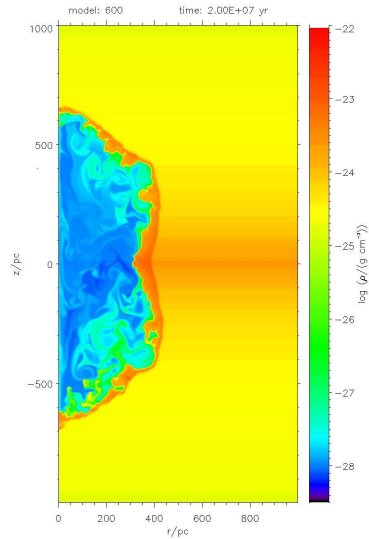} 
\end{center}
\caption{Density distribution of a superbubble after 20 Myrs.
The superbubble results from 100 supernova typeII explosions at the 
origin of the coordinate system.  The temporal sequence of explosions 
happens according to the lifetimes of stars in the mass range between 
10 and 100 $\Msun$ with a Salpeter IMF. 
The galactic disk is vertically composed of the three-phase interstellar 
medium, cool, warm and hot phase, respectively, with 
(central density $\rho_0\, [in\, g\, cm^{-3}]$; 
temperature T [{\it in K}]; 
vertical scaleheight H [{\it in pc}]) of 
($2\times 10^{-24}$; 150; 100), 
($5\times 10^{-25}$; 9000; 1000),
($1.7\times 10^{-27}$; $2\times 10^6$; 4000).
The density varies from almost $10^{-23}\, g\, cm^{-3}$, in the 
densest part of the shell to $2\times 10^{-28}\, g\, cm^{-3}$ in
the darkest bubble interiors. {\it (from \cite{gud02})}
}
   \label{fig1}
\end{figure}
%---------------------------

Such SF trigger by the condensation of swept-up gas in SN or more
efficiently in superbubbles (see fig. 1) can be explored in detail 
by the investigation of the fragmentation timescale 
(see e.g. \cite{ehl97,fuk00}). Ehlerova et al. 
compared a self-similar analytical solution with the results of 3D 
numerical simulations of superbubble expansions in homogeneous media. 
The amount of energy supply from the final number of young stars in an 
OB association, the value of the sound speed, the stratification and 
density of the ambient medium, the galactic differential rotation,
and the vertical gravitational force in the galactic disk, all these 
influence the fragmentation. The typical superbubble radius at which 
shells start to fragment decreases from almost 700 pc at an ambient 
gas density n of 1 cm$^{-3}$ to 200 pc at 10 cm$^{-3}$. 
While in thick disks like they exist in dwarf galaxies (DGs) nearly the 
whole shell fragments, in thin disks it is restricted to the galactic equator only. 
The SF process itself cannot be resolved in these studies and the assumption 
follows the line, that unstable fragments may become molecular and trigger 
the formation of molecular clouds in which new stars are formed. The main 
conclusion is that in DGs the SF may propagate in all directions 
possibly turning the system as the whole into a starburst, while in 
spiral galaxies the SF propagates within a thin strip near the 
symmetry plane only. 
Since the applied thin shell approximation is reasonably only a 
0th-order approximation, in a recent paper \cite{wue10} 
(see also this volume) clarify that the shell thickness and the 
environmental pressure influences the fragments in the sense that 
their sizes become smaller for higher pressure. 
Nevertheless, the deviations from the thin-shell approximations 
are not large. Yet as a drawback the influences of magnetic fields 
is not taken into account.

The perception of SF trigger in SN or superbubble shells sounds
reasonable from the point of view of numerical models because 
(as shown in fig. 1) 
sufficient ambient ISM mass is swept up, preferably in the gas disk 
itself, and is capable to cooling and gravitational instabilities.  
Shell-like distributions of young stars, are e.g. found in
G54.4-0.3, called {\it sharky} (\cite{jun92}), 
in the Orion-Monoceros region (\cite{wil05}), 
more promising in the Orion-Eridanus shell (\cite{lee09}),
and in several superbubbles in the Large Magellanic Cloud,
as e.g. Henize 206 (\cite{gor04}).
Also the formation of Gould's Belt stellar associations in the
shell of a superbubble is most probable (\cite{mor99}), however,
still debated (\cite{com94}).
 
Another possible feedback effect by SN is caused, when the 
ultra-fast SNR shock overruns a dense interstellar cloud, 
so that the clouds are quenched and stars are expected to be formed 
instantaneously. Although such cloud crushing is numerically modelled
its effect on the necessary collapse of sub-clumps is not resolved
simultaneously (\cite{orl05}). 

Since in several DGs excessively high SFRs are observed 
(\cite{hun98,zee98,sti02} and further more), it is a matter of study 
whether starbursts are the result of a SF self-trigger mechanism or the 
issue of an external process because the objects are obviously linked 
to large enveloping gas reservoirs from which gas infall must be 
invoked (\cite{mue05,hen04}). How the hot SN gas that transits
from superbubbles to galactic winds interacts with infalling cold gas 
is yet unexplored but an attractive challenge from the present to
the early universe (\cite{rec07}). High-velocity clouds (HVCs) in 
our Milky Way on their passage through the hot halo gas are expected
to be easily disrupted due to Kelvin-Helmholtz (KH) instability.
In contrast, model clouds of self-gravitating clouds including the effect of 
saturated heat conduction survive over almost 100 Myrs because they 
are mostly stabilized against KH instability (\cite{vie07}). 
While their compression in the stratified halo gas should not be
able to avoid SF, its absence in HVCs is still a mistery, so that
its understanding would provide a further insight into the 
positive vs. negative feedback effect of SN gas on SF in clouds.

\subsection{Supernova energy impact}

Because of their enormous power in various forms, the majority of galaxy 
evolution models usually take the energy deposit of SNeII explosions 
into account as the only heating source for the ISM. 
Although it is generally agreed that the explosive energy of an individual
SN lies around 10$^{51}$ ergs with significant uncertainties (or an 
intrinsic scatter), however, of one order of magnitude, the energy 
deposit as turbulent and consequently as thermal energy to the ISM
is still more than unclear, but it is one of the most important ingredients 
for our understanding of galaxy formation (e.g. \cite{efs00,sil03}). 

As a similar study, the energy release of massive stars as
radiation-driven and wind-blown \HII regions can be considered. 
Analytical estimates for purely radiative \HII regions yielded an energy
transfer efficiency $\e$ of the order of a few percent (\cite{las67}). 
Although the additional stellar wind power $L_w$ can be easily evaluated 
from models and observations, its fraction that is transferred into 
thermal and turbulent energy is not obvious from first principles. 
Transfer efficiencies for both radiative and kinetic energies remain 
much lower from detailed numerical simulations than analytically derived
and amount to only a few per mil (\cite{hen07}, and references 
therein). Nevertheless, as valid for \HII regions and because massive stars 
do not disperse from the SF site, SNeII explode within the stellar associations. 
While massive stars contribute significantly to the ISM structure formation 
as e.g. by cavities and holes,  in the \HI gas and chimneys of hot gas, 
on large scales the energy release by massive stars triggers the matter
circulations via galactic outflows from a gaseous disk and galactic winds.
By this, also the chemical evolution is affected thru the loss of 
metal-enriched gas from a galaxy (for observations see e.g. \cite{mar02},
for models e.g. \cite{rec06b}). 

SN explosions as an immediate consequence of SF stir up the ISM by the 
expansion of hot bubbles, deposit turbulent energy into the ISM, thereby, 
heat the ISM and regulate the SF again (\cite{hen02}). 
This negative energy feedback is enhanced at low gravitation because 
the SN energy exceeds easily the galactic binding energy and drives 
a galactic wind. 
Although investigations have been performed for the heating 
(or energy transfer) efficiency $\eSN$ of SNe (\cite{tho98}), superbubbles 
(e.g. \cite{str04}), and starbursts (\cite{mel04}) they are yet too 
simplistic and mostly spatially poorly resolved to account for quantitative 
results. Thornton et al. derived an efficiency $\eSN$ of 0.1 from 1D 
SN simulations as already applied by chemo-dynamical galaxy models 
(\cite{sam97}), while unity is also used in some galaxy models 
(see sect. 3), but seems far too large.

%---------------------------
\begin{figure}[ht]
\begin{center}
\includegraphics[width=12cm]{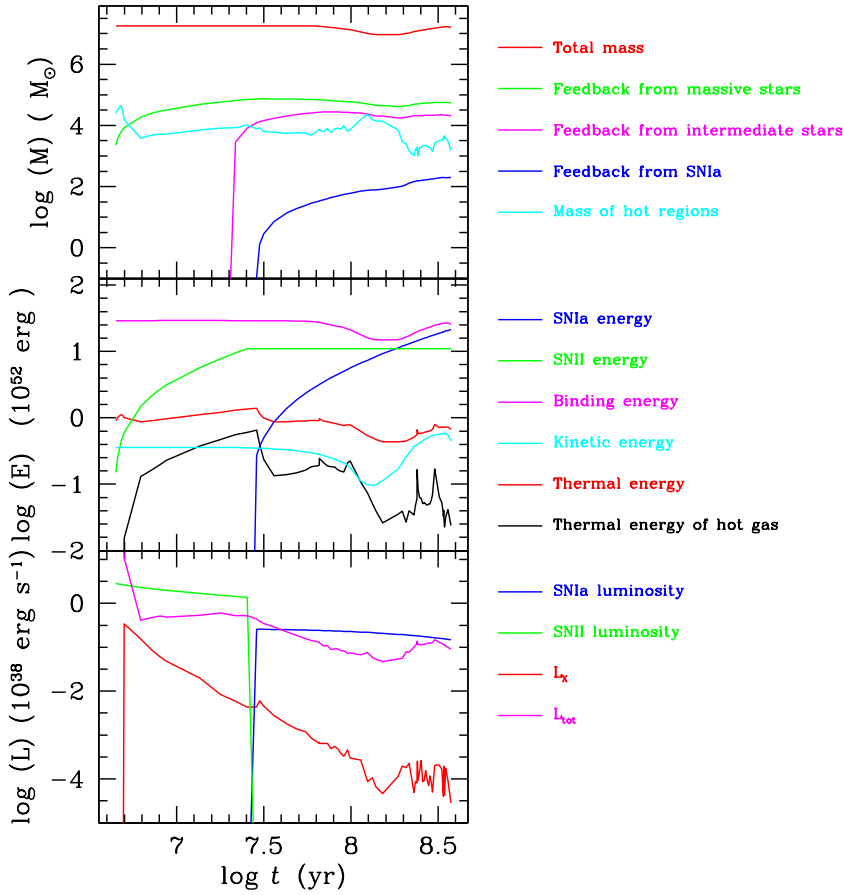} 
\end{center}
\caption{Temporal evolution of masses energies, and luminosities,
respectively, of contained and released components for a single 
star-formation burst in a dwarf galaxy like I~Zw~18 (\cite{rec02}).
The green (light) curves refer to the supernovae typeII contribution.
Comparison of the thermal energy (red line) and the accumulated
supernova typeII energy (middle panel) shows that the thermal energy 
is almost not enhanced by SNeII.
}
   \label{fig2}
\end{figure}
%---------------------------

Although numerical experiments of superbubbles and galactic winds 
are performed, yet they only demonstrate the destructive effect 
on the surrounding ISM but lack of self-consistency and a complex treatment. 
Simulations of the chemical evolution of starburst DGs by 
\cite{rec02}, 2006), that are dedicated to reproduce the peculiar abundance 
patterns in these galaxies by different SF episodes, found, that $\eSN$ 
can vary widely. 
A superbubble expanding from a stellar association embedded in a thick 
HI disk has, at first, to act against the surrounding medium, by this,
is cooling due to its pressure work and radiation, but compresses the 
swept-up shell material and implies turbulent energy to the ISM. 
Here the superbubble expansion is efficiently hampered, but depends 
on the HI disk thickness (\cite{rec09}) and the energy loss 
by radiative cooling. 

In fig.2 the temporal behaviour of the above-mentioned starburst DG model by
\cite{rec02}) is displayed. Comparison of the various kinds of energy 
contents reveal an heating efficiency $\eSN$ of to about 18\%.
While the accumulated SNII energy release reaches $10^{53}$ ergs after 
20 Myrs, the thermal energy content starts at $10^{52}$ drops and increases 
successively but not above $1.8\times 10^{52}$ ergs. For the subsequent 
SNIa explosions, always single events, the accumulated energy release is 
clearly discernible while the thermal energy decreases again and varies 
below $10^{52}$ ergs so that here $\eSN \rightarrow 0$.
Moreover, if a closely following SF episode pushes its SNeII into the 
already existing chimney of a preceding superbubble, the hot gas can 
easily escape 
without any hindrance and thus affects the ISM energy budget much less.
\cite{rec06a} found that depending on the external \HI density the 
chimneys do not close before a few hundred Myrs.

\section{Galaxy evolution}

Numerical simulations of galaxy evolution are only feasible with 
intermediate-scale spatial resolution so that SF cannot be resolved 
and must already be prescribed by reasonable recipes. 
Numerous papers have implemented SF criteria, such as e.g. 
threshold density $\rSF$ with SF if $\rg \ge \rSF$, 
excess mass in a specified volume with respect to the Jeans mass, 
i.e. M$_{\rm g} \ge M_{\rm J}$, 
convergence of gas flows ($div\cdot \underline{\bf v} < 0$), 
cooling timescale $t_{cool} \le t_{dyn}$, 
temperature limits T$_{\rm g} \le$ T$_{lim}$, 
Toomre's Q parameter, 
temperature dependent SFR,
but to some extent also already under the assumption of the KS law 
(\cite{dvs08}). This latter, however, is unjustified for
non-equilibrium situations, because the SF should converge to this
relation due to self-regulation. 
If at least some of these conditions are fulfilled, as a further step,
the gas mass which is converted into stars, i.e. the SFE, has to be set 
e.g. by fixing an empirical value $\eSF$ from observations, 
i.e. $\Delta m_{\rm SF} = \eSF \cdot \rg \cdot \Delta x^3$, where 
$\Delta x^3$ is e.g. the mesh volume in a grid code.
If the numerical timestep $\Delta t$ is smaller than the dynamical 
timestep $\tff$, $\Delta m_{\rm SF}$ has to be weighted by this time 
ratio (\cite{tas06}). Since $\eSF$ must inherently depend on the 
local conditions so that it is high in bursting SF modes, as requested 
for the Globular Cluster formation, but of percentage level in the 
self-regulated SF mode, numerical simulations often try to derive the 
realistic SFE by comparing models of largely different $\eSF$ with 
observations, as e.g. to reproduce gas structures in galaxy disks and
galactic winds (e.g. $\eSF$ = 0.05 and 0.5 in \cite{tas06}, 2008). 
In addition, $\eSN$ by them is fixed to 
10$^{51}$ erg per 55 $\Msun$ of formed stars  
($1.8\times 10^{49}$ erg $\Msun^{-1}$ by \cite{dvs08}), 
so that these results can be treated indicatively but not yet 
quantitatively, since they also mismatch with the KS relation.

Theoretical studies by \cite{elm97} achieved a dependence of $\eSF$ on the
external pressure, while \cite{koe95} explored a temperature dependence of 
the SFR. Furthermore, most galaxy evolutionary models at present lack 
of the appropriate representation of the different ISM phases allowing for
their dynamics and their
direct interactions by heat conduction, dynamical drag, and dynamical 
instabilities thru forming interfaces, not to mention resolving the
turbulence cascade.

\section{Conclusions}

The dominating influence of SN explosions and superbubbles on structure,
dynamics, and energy budget of the ISM are obvious and agreed.
Signs and strengths of these feedback effects are, however, widely 
uncertain. Whether the feedback is positive (trigger) or negative 
(suppression) can be understood analyticly from first principles,
but because of the non-linearity and the complexity level of the 
acting plasmaphysical processes clear results cannot be quantified 
reliably. In addition, the temporal behaviour varies by orders of 
magnitude because of the changing conditions.
In summary, the energy transfer efficiencies of SNe and superbubbles 
to the ISM are much below unity and depend on the temporal 
and local conditions, but must not be overestimated.
Spatially and temporarily resolved simulations of SN and superbubbles
in an extended environment with varying conditions of the ISM are 
necessary in order to connect large-scale effects on SF clouds 
with the existing detailed simulations on the star-forming scales.

\begin{acknowledgments}
The author is grateful to Simone Recchi for substantial contributions 
to this topic and providing fig. 2.
The attendance of the symposium was funded by the key programme 
''Computational Astrophysics'' of the University of Vienna under project 
no. FS538001.
\end{acknowledgments}

\end{document}